# Income Inequality, Cause and Cure

B.N. Kausik[1]

## Abstract

We argue that the recent growth in income inequality is driven by disparate growth in investment income rather than by disparate growth in wages. Specifically, we present evidence that real wages are flat across a range of professions, doctors, software engineers, auto mechanics and cashiers, while stock ownership favors higher education and income levels. Artificial Intelligence and automation allocate an increased share of job tasks towards capital and away from labor. The rewards of automation accrue to capital, and are reflected in the growth of the stock market with several companies now valued in the trillions. We propose a Deferred Investment Payroll plan to enable all workers to participate in the rewards of automation and analyze the performance of such a plan.

---

[1] Conflict disclosure: unaffiliated independent, previously entrepreneur, engineer and academic.
Author's bio: https://www.linkedin.com/in/bnkausik/ Contact: bnkausik@gmail.com

# Introduction

Fig. 1 compares the real median household income and real median GDP per household in the US between 1984 and 2020 as compiled from FRED Census Bureau data by the St. Louis Fed. Household GDP grew at a CAGR of 1.2%, while household income grew at 0.6%. In short, productivity grew twice as fast as median income.

**Fig. 1: Normalized Median Income & GDP**

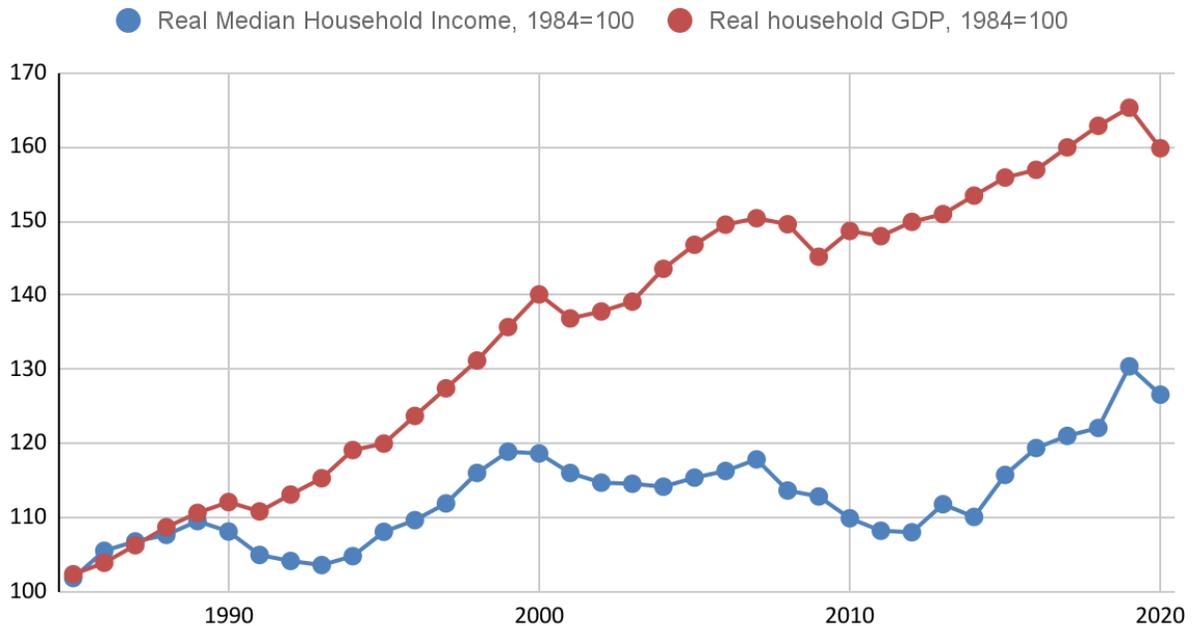

Source: fred.stlouisfed.org

The productivity gains benefited the top income bands, at the cost of the bottom, as shown in Fig. 2 below from Horowitz, Igielnik and Kochar (2020). Specifically, the ratio of the income of the top 10% to the bottom 10% increased from 9.1 in 1980 to 12.6 in 2018.

**Fig. 2: US 90/10 income ratio**

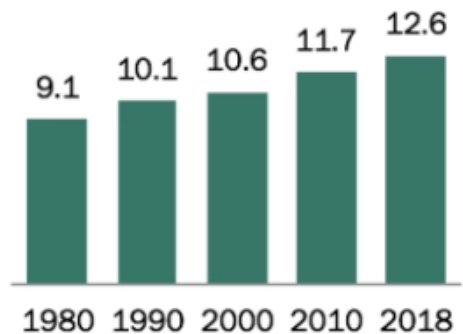

Source: pewresearch

Against the backdrop of growing GDP combined with rising income inequality, economists are broadly agreed on the decline of the labor's share of economic output in industrialized nations, Blanchard (1997), Karabarbounis and Neiman (2013), Piketty (2014) and Dao et al. (2017). Some of the decline is due to the movement of work offshore, Elsby, Hobijn, and Sahin (2013). However, Autor et al. (2020), observe the decline in labor's share even in industries such as wholesale, retail, and utilities, where the opportunity for offshore labor is limited. Furthermore, the decline in labor share affects most industrialized economies, as noted in Autor et al. (2020) and studied at the plant-level by Bockerman and Maliranta (2012).

Several economists have proposed remedies involving the centralized regulation of innovation in automation, particularly of Artificial Intelligence technologies. Korinek and Stiglitz (2020) write:

> *Rapid progress in new technologies such as Artificial Intelligence has recently led to widespread anxiety about potential job losses. This paper asks how to guide innovative efforts so as to increase labor demand and create better-paying jobs. We develop a theoretical framework to identify the properties that make an innovation desirable from the perspective of workers, including its technological complementarity to labor, the factor share of labor in producing the goods involved, and the relative income of the affected workers. Examples of labor-friendly innovations are intelligent assistants who enhance the productivity of human workers. The paper also discusses measures to steer technological progress in a desirable direction for workers, ranging from nudges for entrepreneurs to changes in tax, labor market and intellectual property policies to direct subsidies and taxes on innovation. In the future, we find that progress should increasingly be steered to provide workers with utility from the non-monetary aspects of their jobs.*

Brynjolfsson (2022) writes:

> *In 1950, Alan Turing proposed an "imitation game" as the ultimate test of whether a machine was intelligent: could a machine imitate a human so well that its answers to questions are indistinguishable from those of a human. Ever since, creating intelligence that matches human intelligence has implicitly or explicitly been the goal of thousands of researchers, engineers and entrepreneurs. The benefits of human-like artificial intelligence (HLAI) include soaring productivity, increased leisure, and perhaps most profoundly, a better understanding of our own minds. But not all types of AI are human-like—in fact, many of the most powerful systems are very different from humans —and an excessive focus on developing and deploying HLAI can lead us into a trap. As machines become better substitutes for human labor, workers lose economic and political bargaining power and become increasingly dependent on those who control the technology. In contrast, when AI is focused on augmenting humans rather than mimicking them, then humans retain the power to insist on a share of the value created. What's more, augmentation creates new capabilities and new products and services, ultimately generating far more value than merely human-like AI. While both types of AI can be enormously beneficial, there are currently excess incentives for automation rather than augmentation among technologists, business executives, and policymakers.*

And Acemoglu (2022) writes:

> *"This essay discusses several potential economic, political and social costs of the current path of AI technologies. I argue that if AI continues to be deployed along its current trajectory and remains unregulated, it may produce various social, economic and political harms. These include: damaging competition, consumer privacy and consumer choice; excessively automating work, fueling inequality,* *inefficiently pushing down wages, and failing to improve worker productivity; and damaging political discourse, democracy's most fundamental lifeblood."*

Centralized regulation of AI automation will simply cause those job tasks to move offshore. Furthermore, with global fertility rates continuing to plummet below the replacement rate as shown in Fig. 3, increased automation is an urgent imperative for the future well-being of humans, rather than something to minimize via the centralized regulation of innovation. Specifically, as shown in Fig. 4, the median age of the global population is projected to cross 40 by the year 2100, with fewer younger people available to care for more older people. Such is already the case in countries where the fertility rate is far below the replacement rate.

**Fig. 3: Global Fertility Rate**

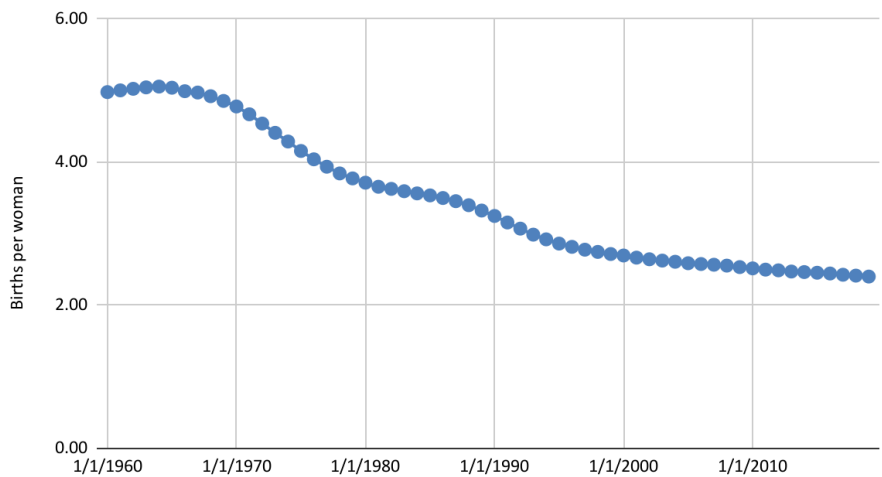

Source: fred.stlouisfed.org

**Fig. 4: Median Age of Global Population**

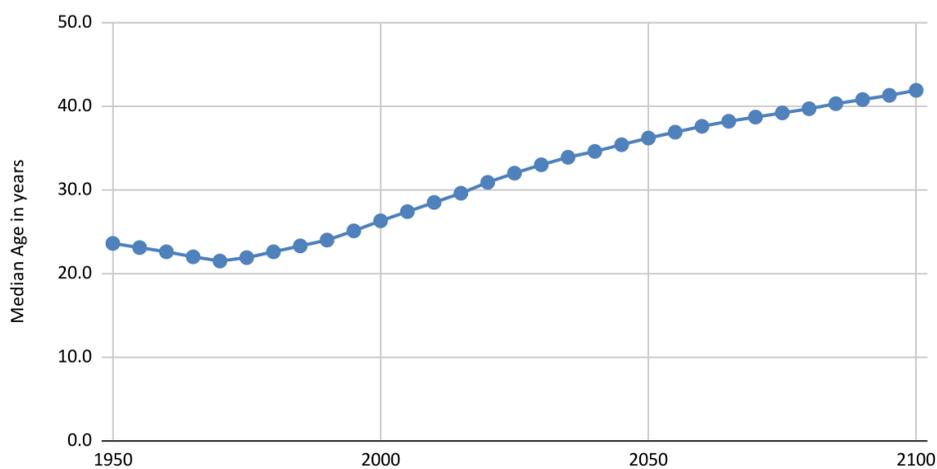

Source: United Nations

From a technologist's perspective, automation in the Information Economy is no different from automation in the Industrial Economy. In the Industrial Economy, physical tasks were automated, e.g. electric drills and metal presses replaced hand drills and anvils. This freed up humans from having to perform repeated and physically exhausting tasks in the mass production of automobiles or household appliances. But humans still had to design and build the electric drills and presses, since these machines could not, and still cannot, build themselves. The same is true of the Information Economy that drives GDP growth in recent years. Computerized automation replaces humans at repeated and mind-numbing tasks, e.g a spreadsheet is vastly preferable to a paper ledger for accounting. But spreadsheet programs cannot write themselves, and increasingly carry AI capabilities like auto-correct and auto-fill. Likewise, an automated cashier at a grocery store frees up human cashiers in the same way that automated phone exchanges freed up telephone operators who manually connected the caller with the recipient. In both cases, the automation technology made it possible for the intent of the customer to be simply and directly translated into the desired outcome without the need to involve another human for action.

Acemoglu and Restrepo (2020) write:

> *"Automation technologies expand the set of tasks performed by capital, displacing certain worker groups from employment opportunities for which they have comparative advantage."*

In short, increasing automation improves human productivity, freeing up humans from repetitive tasks that are both physically and mentally exhausting. The rewards and benefits of automation accrue to the stockholders of corporations investing in or producing automation technologies. Rather than centrally regulating innovation, we propose economic policies to ensure broad stock ownership so that the benefits of automation accrue to all households, rather than just households with higher levels of education and income. In what follows, we distinguish between wages and income as below, from Google.

> *wages*
> *a fixed regular payment, typically paid on a daily or weekly basis, made by an employer to an employee, especially to a manual or unskilled worker*
> *income*
> *money received, especially on a regular basis, for work or through investments.*

Wages are compensation for labor. Income is summed over all sources, including wages and investment income from stock and other vehicles. The Census Bureau does not include capital gains or non-cash benefits such as food stamps in their definition of income.

> *Census money income is defined as income received on a regular basis (exclusive of certain money receipts such as capital gains) before payments for personal income taxes, social security, union dues, medicare deductions, etc. Therefore, money income does not reflect the fact that some families receive part of their income in the form of noncash benefits, such as food stamps, health benefits, subsidized housing, and goods produced and consumed on the farm. In addition, money income does not reflect the fact that noncash benefits are also received by some nonfarm residents which may take the form of the use of business transportation and facilities, full or partial payments by business for retirement programs, medical and educational expenses, etc.*

# Wage Stagnation

We analyze wages across four professions, (1) cashiers, (2) auto mechanics, (3) software developers and (4) physicians from the period 2010 to 2020 via data available at the US Bureau of Labor Statistics, using the internet archive at web.archive.org to retrieve historical data. See Fig. 5.

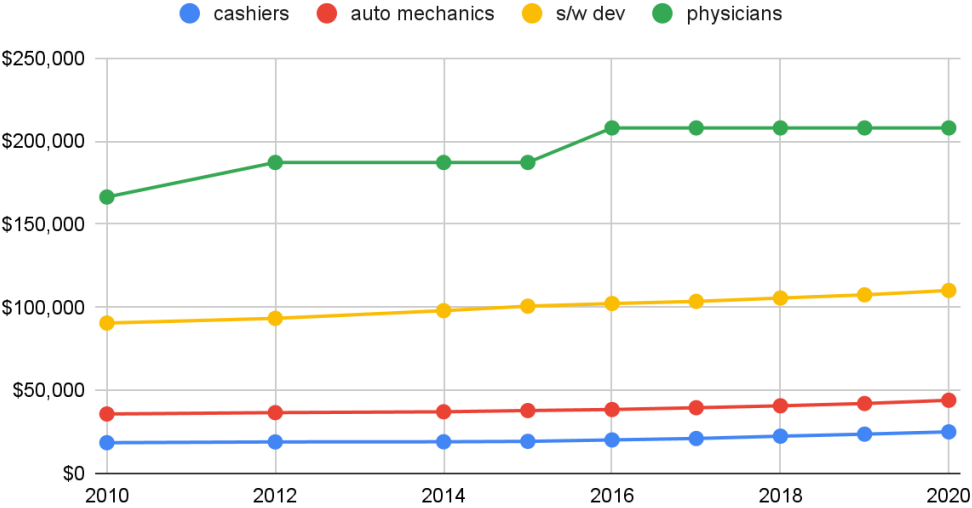

**Fig. 5: Median Wages** (not adjusted for inflation)

Source: bls.gov

Both Software Developers and Physicians typically require bachelor's or graduate degrees, while Cashiers and Auto-mechanics typically have some high school education followed by on the job training. To examine wage growth, we normalize the wages and adjust for CPI inflation as shown in Fig. 6.

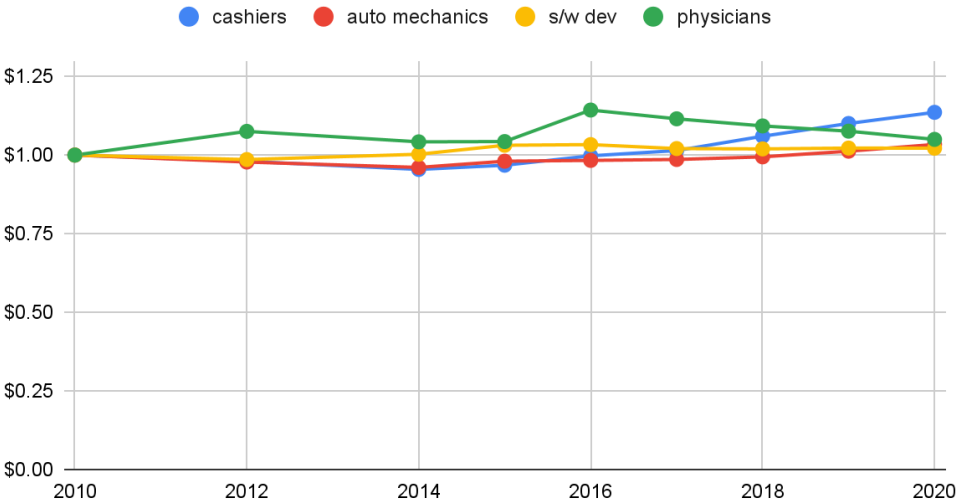

**Fig. 6: Normalized Real Wages** (2010 = 1)

Source: bls.gov

As evident in Fig. 6, none of the professions enjoyed significant wage growth during the period 2010 to 2020.  If anything, Cashiers enjoyed the most wage growth, probably due to raises in minimum wage in some states.

**Fig. 7: Physician Wage Growth** (1970 to 2020)

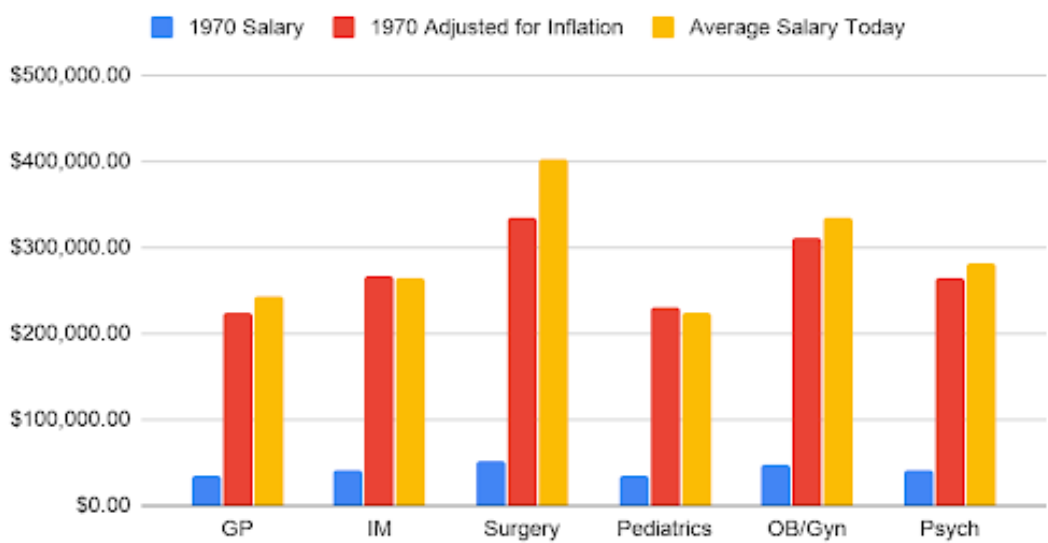

Source: mdlinx

Fig. 7 expands on the time span to show that real wages for physicians were largely unchanged between 1970 and 2020.

In contrast to our findings above, Autor (2019) states:

> *"One of the enduring paradoxes that has accompanied the rise of wage inequality over the last four decades in industrialized economies is the sustained fall in real wages experienced by less-educated workers"*

However, Autor (2019) analyzes men and women separately, and notes that while the least educated men lost wages in real terms, the same was not true for women. Furthermore, Autor (2019) does not distinguish between income and wages, using income data from the Census Bureau's "March Current Population Survey Annual Social and Economic Supplement data."  We treat income and wages as distinct, wages being one component of income.

In summary, wages are broadly stagnant irrespective of education or profession, even in the case of jobs such as physicians and auto mechanics that are difficult to automate or move offshore.

# Income Growth

Income is summed over all sources, including wages and investment income.  With wages broadly stagnant across professions and education levels, we now turn to investment income, particularly the performance of stock. Fig. 8 shows the ten fold inflation adjusted growth of the S&P 500 index over the period Jan 1, 1985 to Dec 31, 2021. The benefits of automation accrue to shareholders of corporations, which in turn is reflected in the growing stock price of the 500 industry leaders comprising the index. Many of these stocks pay a quarterly dividend to shareholders, resulting in additional income.

**Fig. 8: S&P 500 Adjusted for Inflation (1985 to 2022)**

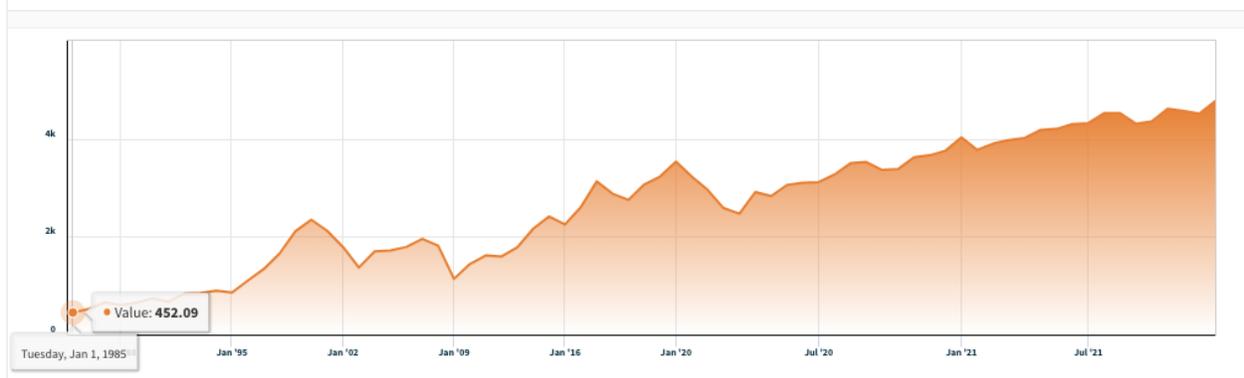

Source: Nasdaq

Stock ownership is tilted in favor of households with higher education levels, Hryshko et al. (2012), and in favor of higher incomes, Parker and Fry (2020). Fig. 9 shows that families with incomes over $100,000 per year own stocks at 4X the rate as families with incomes below $35,000 per year.

**Fig. 9: Stock Ownership by Families**

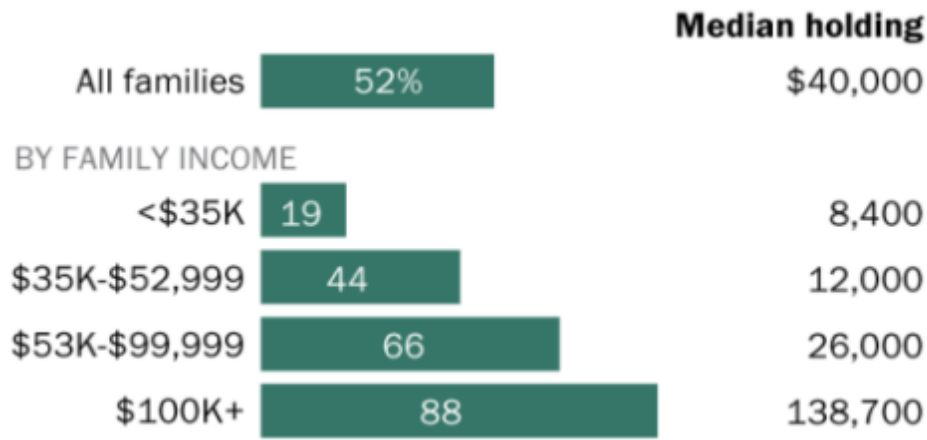

Source: pew research

In brief, the financial benefits of automation accrue to capital rather than to labor, i.e. stockholders of the corporation. Since stock ownership is tilted towards the educated and higher income households, the income derived from automation accrues disproportionately to educated and higher income households.

# Reducing Income Inequality

In technology companies, it is common for labor to participate in the stock gains of the employer via stock options and restricted stock units. Technologists produce automation and understand well that stock ownership is the path to prosperity in the face of automation. It is no accident that the Forbes 400 list of the wealthiest Americans includes many technology entrepreneurs, who represent both capital - in that they are significant stockholders in companies that favor automation, and labor - in that they earn their salary and stock ownership as compensation for their work. Blasi et al. (2008), Bryson et al. (2016) consider the pros and cons of the "shared capital" model where employees participate in their employer's stock.

We propose a Deferred Investment Payroll (DIP) plan for all workers across the economy, invested in a broad stock market index fund. In our proposal, employees receive a percentage of their wages deferred for a period during which the money is invested in the broad stock market. Money is paid in and paid out in a rolling fashion, and the payout includes both appreciation in principal as well as ongoing dividends. At retirement, all accumulated holdings are paid out as a lump sum. Specifically, we consider the following example structure:

**Deferral Period:** $t = 1, 2, 3, ..., T$
**Deferral Amount in year** $t$**:** employer contributes percentage $c$ of wages into investment

**Deferral Investment:** S&P 500 index fund
**Deferral Payout in year** $t$**:**
　　Dividends: paid as received
　　Principal: percentage $p_t$ of investment liquidated and paid out
**Lump sum payout at end of plan:** all investment liquidated and paid out

We first examine the performance of the plan on a model over a plan term $T = 45$ years for a hypothetical employee who tracks the median income. The model reflects the aggregate averages of historical data between 1985 and 2020.

**Real Household GDP:** 1.2% CAGR
**Real Median Income:** 0.6% CAGR
**Inflation:** 2.6% per year
**S&P 500:** 8.8% CAGR (not adjusted for inflation)
**S&P 500 Dividends:** 2.3% per year

We have several factors to balance
- Close the gap between median income and GDP growth.
- Minimize employer contribution
- Zero residual balance at end of plan term

Appendix A derives the optimal values of $c$ and $p_t$ as the solution to a multivariate system of nonlinear equations. Specifically, setting $p_T = 1$ to ensure zero residual balance, working backwards for $p_{T-1}, P_{T-2}, ... p_1$ and then iterating over $c$ to minimize the deviation between normalized median income and normalized GDP gives us $c = 0.0872$ or 8.72%.

Fig. 10 shows the impact of DIP for the model at the optimal values, compared against a simple pay raise without deferred investment. In the figure, income from DIP almost exactly overlays the normalized real GDP.

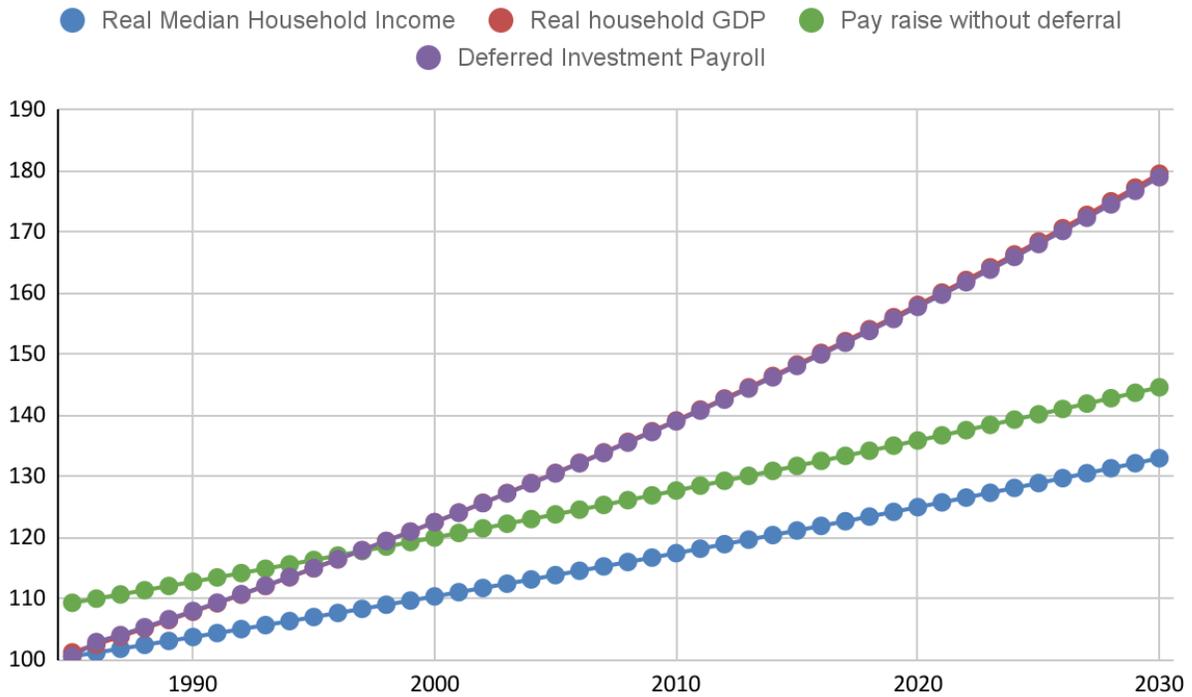

**Fig. 10: Normalized Model Impact** (1984 dollars)

In the context of the model, DIP outperforms a simple pay raise after an initial ramp period, closing the gap between productivity and income growth. Over the course of the plan term of 45 years, for the first 15 years plan inflow exceeds outflow, while for the remaining 30 years plan outflow exceeds inflow.

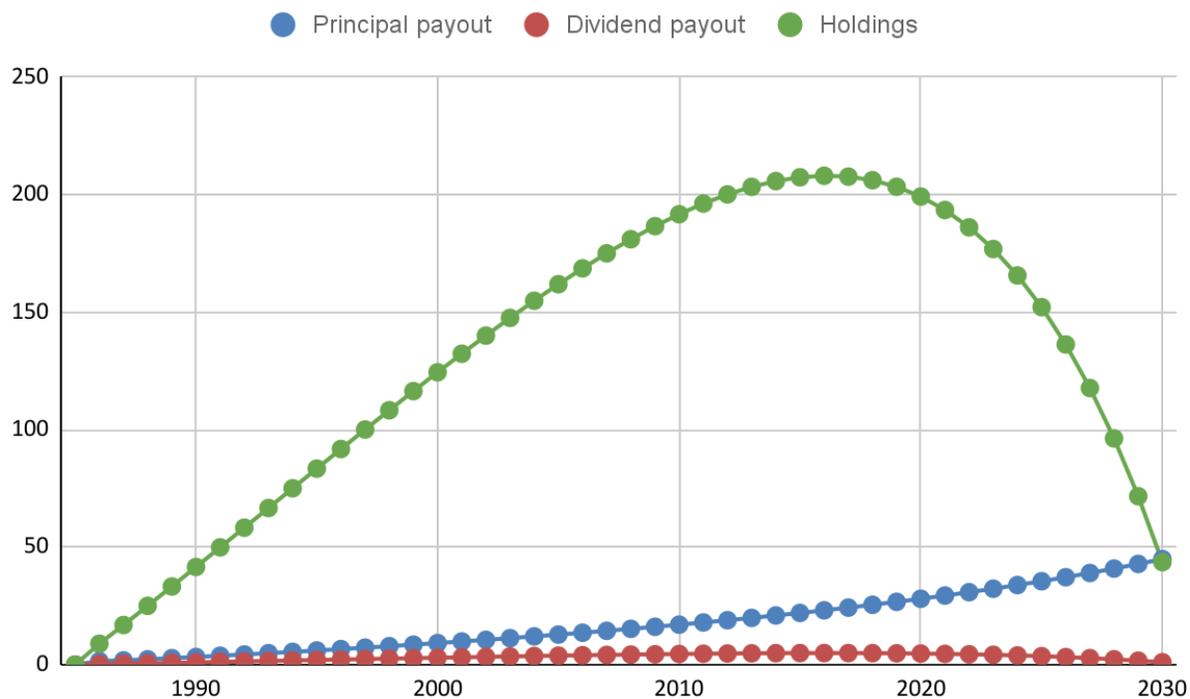

Fig. 11 shows the principal and dividends paid out, as well as the accumulated holdings over the plan period. The average of the accumulated holdings is approximately the annual median household income, and there are no residual funds at the end of the plan term. Fig. 12 shows the historical performance of DIP between 1985 and 2020, using the same choices for $c$ and $p_t$, as well as a plan term of 45 years.

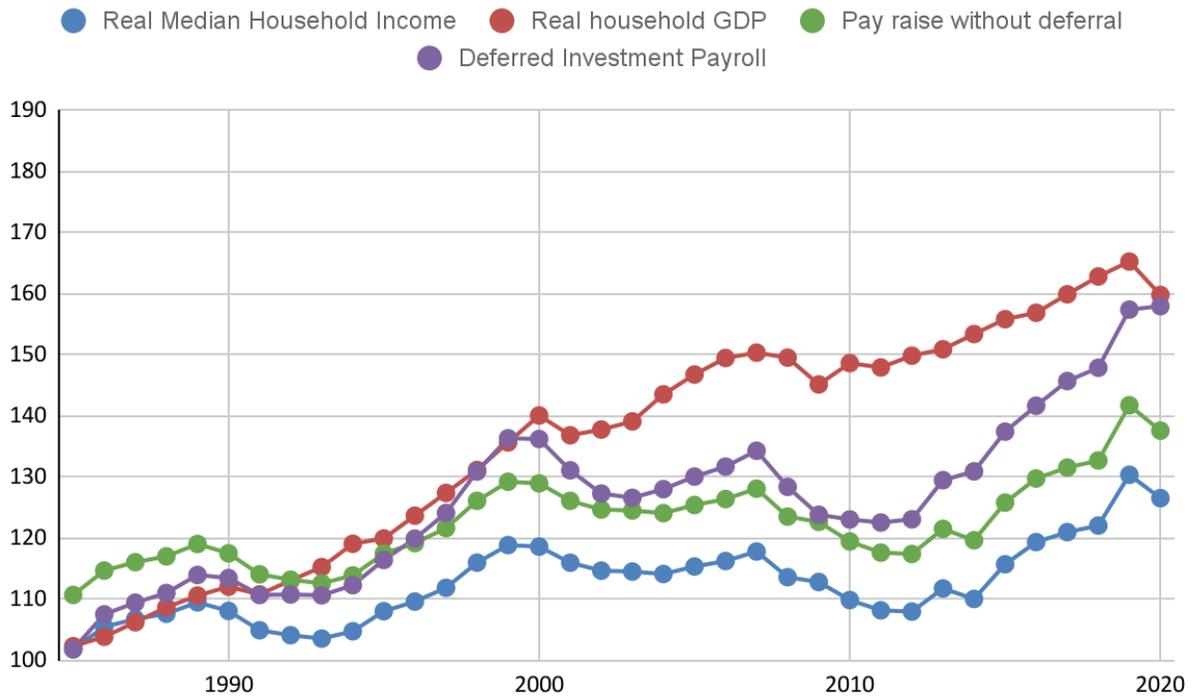

Fig. 12: Normalized Historical Impact (1984 dollars)

Sources: Inflation: minneapolisfed.org; S&P 500: macrotrends.net; S&P 500 dividends: stern.nyu.edu

Fig. 13 shows the principal and dividends paid out, as well as the accumulated holdings over the plan period. The holdings in 2020 do not go to zero since the plan term runs to the year 2030.

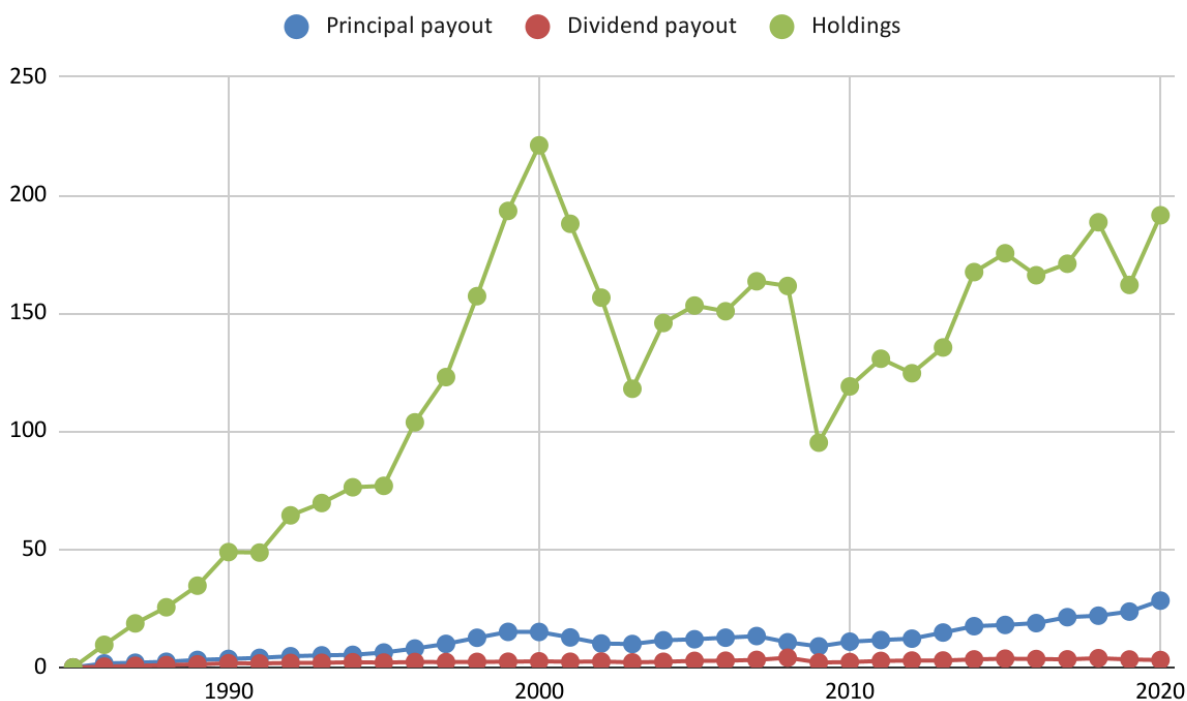

**Fig. 13: Normalized Historical Funds Flow** (1984 dollars)

It is clear that DIP boosts the median wage by accruing the rewards of automation to all workers, rather than just educated and upper income households.

DIP raises a number of policy questions, some examples below:

- Choosing model parameters that retain principal at end of plan can boost retirement savings.
- Capping the contribution as a percentage of the median income with annual adjustments would benefit the lower income levels while minimizing market impact and competition. In our current model, the average holdings over the term of the plan is approximately the median household income. At $44,000 per employee across the ~150M non-farm workforce in December 2021 per bls.gov, the total holdings at equilibrium would be $6.6T. As of January 2022, the total capitalization of the S&P 500 is ~$40T, and of the broader US stock market over $50T.
- Second, DIP blurs the distinction between capital gains from investments and regular income from wages. This benefits lower income households who otherwise do not participate in the tax structure that favors capital gains, but requires a recalibration of census income data collection.
- Third, it is logical to consider excluding government employees from the plan, as they have other protections and guarantees.

14# Summary

Over the past forty years, real median household income in the US has been essentially flat despite growth in productivity as reflected in real median household GDP. There is growing agreement that Artificial Intelligence automation has allocated a growing share of job tasks towards capital, and away from labor. The rewards of increased automation accrue to capital, and are thereby reflected in the growth of the stock market with several companies now valued in the trillions. Noting that household income includes both wages and investment income from stocks, we presented evidence that while wages are flat across a broad range of income levels and education, stock ownership is tilted to households with higher income and education levels, resulting in uneven apportionment of the rewards of automation. We proposed a Deferred Investment Payroll system where all workers benefit from stock ownership. We analyzed how such a policy bridges the gap between productivity and median income on model and historical data, thereby reducing income inequality. Should Deferred Investment Payroll be mandated by policy, it can be implemented by financial institutions and payroll processors similar to IRA and 401(k) programs.# Acknowledgements

Thanks to Forest Baskett, Gopal Sharathchandra, Jay Jawahar, Lawrence Katz and Prasad Tadepalli for their comments and suggestions.

# Appendix A

In this section, we optimal values of $c$ and $p_t$ for the model. For year $t = 1, 2, 3..., T$, let $I_t$ be the normalized real median income; $G_t$ be the normalized real household GDP and $H_t$ the real stock holdings. Let $s$ be the real CAGR of the stock holdings, and $d$ the annual dividend rate. By definition, we have

$$H_{t+1} = (H_t(1 - p_t) + cI_t)(1 + s) \quad (1)$$

The total income in year $t$ is

$$H_t(p_t + d) + I_t$$

To close the gap between median income and GDP, we would like the total income to be

$$H_t(p_t + d) + I_t = G_t$$

Rearranging, we get

$$H_t = (G_t - I_t)/(p_t + d) \quad (2)$$

Substituting Equation (2) in (1) and rearranging, we get

$$\left(\frac{1-p_t}{p_t+d}\right) = \left(\frac{G_{t+1} - I_{t+1}}{(p_{t+1}+d)(1+s)} - cI_t\right)\left(\frac{1}{G_t - I_t}\right) \quad (3)$$

Setting $\alpha_t$ to be the right-hand side of Equation (3) and solving, we get

$$p_t = (1 - d\alpha_t)/(1 + \alpha_t) \quad (4)$$

Since we want the residual stock holdings at the end of the term to be zero, we set $p_T = 1$. Substituting in Equation (4), we can compute $p_{T-1}, p_{T-2}, .... p_1$ for a given value of $c$. For each value of $c$, let $\hat{p}_t = max(p_t, 0)$, the total income $\hat{I}_t = I_t + (\hat{p}_t + d)H_t$, and the loss function

$$\sum_{t=1}^{T} |1 - \hat{I}_t/G_t| \quad (5)$$

Iterating over values of $c$ to minimize the loss function we find the optimum at $c = {\sim}0.0876$ or 8.76%, see Fig. 14.

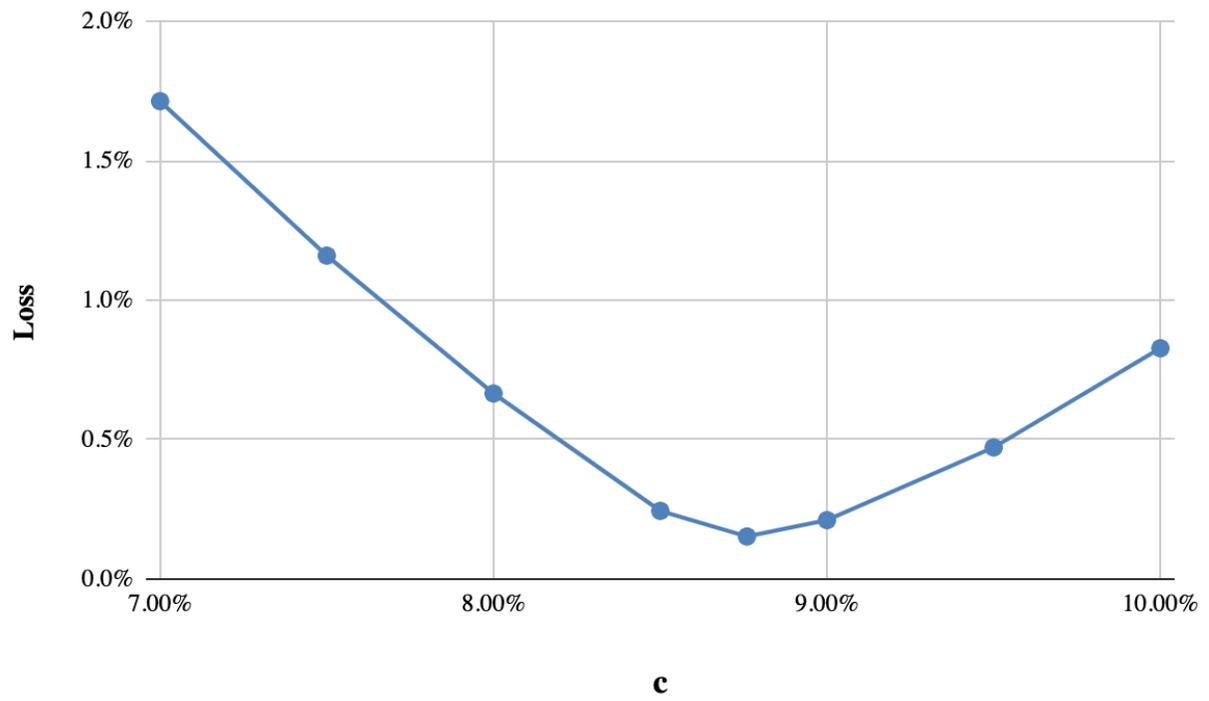

**Fig. 14: Optimum Value of Model Contribution $c$**

18Parker, Kim and Fry, Richard, (2020): More than half of U.S. households have some investment in the stock market.
https://www.pewresearch.org/fact-tank/2020/03/25/more-than-half-of-u-s-households-have-some-investment-in-the-stock-market/

Blasi, J. R., Freeman, R. B., Mackin, C. and Kruse, D. L. (2008) "Creating a Bigger Pie? The Effects of Employee Ownership, Profit Sharing, and Stock Options on Workplace Performance", NBER Working Paper No. 14230

Bryson, A., Clark, A. E., Freeman, R.B., and Green, C. P. (2016) "Share Capitalism and Worker Wellbeing", Labour Economics, 42: 151-158